\documentclass[prb,aps,twocolumn,epsfig]{revtex4}

\def\pr{Phys. Rev.\ }

\def\tit#1#2#3#4#5{{#1} {\bf #2}, #3 (#4)}

\def\prl{Phys.\ Rev.\ Lett.\ }
\def\pr{Phys.\ Rev.\ }
\def\prb{Phys.\ Rev.\ B\ }

\def\jpa{J.\ Phys.\ A\ }

\def\natu{Nature\ }
\def\jsp{J.\ Stat.\ Phys.\ }

\def\jmp{J. Math.\ Phys.\ }

\newcommand{\bea}{\begin{eqnarray}}
\newcommand{\eea}{\end{eqnarray}}

\usepackage{epsfig}

\begin{document}
%\twocolumn[\hsize\textwidth\columnwidth\hsize\csname @twocolumnfalse\endcsname
\title{
Pocket Monte Carlo algorithm for classical doped dimer models
}

\author{
Werner Krauth$^1$ and R. Moessner$^2$
}
\email{krauth@lps.ens.fr; moessner@lpt.ens.fr}

\affiliation{
$^1$CNRS-Laboratoire de Physique Statistique de l'Ecole Normale
Sup\'erieure\\
$^2$Laboratoire de Physique Th\'eorique de l'Ecole Normale
Sup\'erieure, CNRS-UMR8541\\
24, rue Lhomond, 75231 Paris Cedex 05, France}
%\date{\today}

%\maketitle

\begin{abstract}
We study the correlations of classical hardcore dimer models 
doped with monomers
by
Monte Carlo simulation. We introduce an efficient cluster algorithm,
which is applicable in any dimension, for different lattices and
arbitrary doping.  We use this algorithm for the dimer model on the square
lattice, where a finite density of monomers destroys the critical
confinement of the two-monomer problem. The monomers form a two-component
plasma located in its high-temperature phase, with the Coulomb
interaction screened at finite densities. On the triangular
lattice, a single pair of monomers is not confined. The monomer
correlations are extremely short-ranged and hardly change with doping.

\end{abstract}

%]

\maketitle

\section{Introduction}
Models of classical dimers on a lattice play an important role in
polymer physics, but their statistical mechanics has proven to be of
great value in a much wider range of settings. Prominently, in the
early days of the resonating valence bond theories\cite{Anderson87} 
of high-temperature
superconductivity, Rokhsar and Kivelson\cite{Rokhsar88} proposed that
dimers could act as caricatures of singlet (`valence') bonds between
spins 1/2, with the hardcore condition on the  dimer
configurations translating the constraint that each spin participate
in exactly one singlet bond with a neighboring spin.  Resonances
between different configurations generate a quantum dynamics for the
resulting Rokhsar Kivelson quantum dimer models (RK-QDM).

These RK-QDMs are also of interest as effective theories, as they
appear as limiting cases of $d=2+1$\ dimensional Ising gauge theories
which are dual to frustrated Ising models.\cite{msf} Central in this
context is the question of confinement -- can two test charges be
separated infinitely far at finite cost in free energy? In the dimer
model, these test charges are monomers, representing spinons or holons
in the RVB theory.

In the context of RK quantum dimer models, \emph{classical} dimer models
play a crucial role because, for a particular choice of parameters,
the quantum wavefunction is an equal-amplitude superposition of all
classical dimer configurations.\cite{fn-sectors} For this reason,
dimer-diagonal correlation functions can be obtained from the classical
(equally weighted) average over all dimer configurations. At zero doping,
such correlations can be obtained analytically using Pfaffian methods
introduced by Kasteleyn\cite{kasteleyn} and developed further by a number
of authors.\cite{dimerrev} These methods have been extended to obtain
correlation functions of a pair of monomers on the square
lattice,\cite{Fisher63,Hartwig} 
although it is not clear whether asymptotic
monomer correlations can in general be obtained in closed form
along those lines. Present analytical approaches do not allow an exact
treatment of finite monomer densities; this is related to the absence
of  an analytical solution  of the two-dimensional Ising model in a
magnetic field.\cite{kast2}

To reach the RK point in the presence of doping, it is necessary not
only to fine-tune two ratios of magnitudes of kinetic and potential
energies but also a hopping phase.\cite{Rokhsar88} This procedure is not
innocuous, but classical correlations are also interesting because they
are those of the quantum model at high temperature. There,  hopping and
resonance as well as any potential energies included in the dimer model
are much smaller than the thermal energy.  The classical correlations
can remain non-trivial as a result of the projection onto the space
of monomer-dimer coverings.  One is thus interested in the presence of
(at least short-range\cite{liebhei}) correlated structures, and how they
change with doping.

In this paper, we present an efficient numerical algorithm for simulating
classical monomer-dimer models. The algorithm is analyzed in some detail,
and we mention possible uses for related problems.

Our main results are the following. Firstly, a pair of monomers on the
triangular lattice is deconfined, with a correlation length of less
than one lattice constant. This is in keeping with the very
short-range \emph{dimer} correlations on that lattice.\cite{MStrirvb} On the
square lattice, the critical dimer correlations\cite{Fisher63}
generate a Coulomb interaction, with monomers on the
same (different) sublattice having equal (opposite) charges. At finite
density, the monomers form a two-dimensional two-component plasma in its
Debye-screened high-temperature phase.  For both square and triangular
lattice, the monomer correlations decay monotonically with increasing
density, without any sign of additional new correlations on any
length scale.

We note that the study of monomer-dimer models goes back a long
time.\cite{rushfow} However, we have not found in the literature an
evaluation of the correlations studied here.

\section{Pocket algorithm for Monte Carlo simulation of  dimer models}

\begin{figure}
\centerline{ 
\psfig{figure=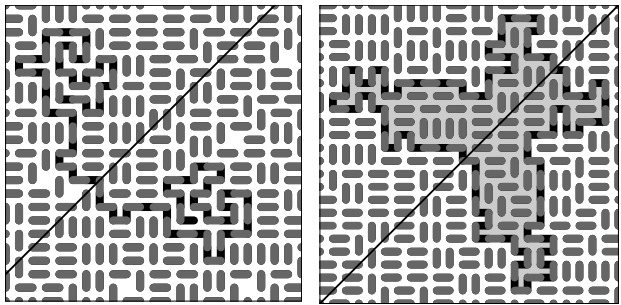,height=4.0cm} }
\caption{Transition graphs  of initial (black) and final (gray) configurations
of the pocket algorithm presented in this paper. 
Left: In the presence of monomers the graph can be open;
right: transition loop touching the symmetry axis.
The cluster move consists in flipping the cluster - 
Note that we may also flip the dimers enclosed by a loop, and that 
the transition graph crosses the symmetry axis at most twice.  }
\label{f:graph_and_loop}
\end{figure}

Overlaying two hardcore dimer configurations generates what is known
as their transition graph; this graph consists of disjoint subgraphs of
dimers alternating between the two configurations -- such subgraphs
are shown in Fig.~\ref{f:graph_and_loop}. 

An open graph, as shown on the left of
Fig.~\ref{f:graph_and_loop}, must terminate on a monomer.  Otherwise,
it is a closed loop (a ``transition graph loop''), as the one
shown on the right of Fig.~\ref{f:graph_and_loop}.

Any  Monte Carlo move corresponds to  a transition graph of the initial
and final configurations.  The simplest transition graphs arise from two
pairs of dimers (for the square and triangular lattice) or from three
(hexagonal lattice), which wind around a unit cell of the
lattice. A Monte Carlo algorithm based on these local moves can have
the same convergence problems as the local simulation methods for, e.g.,
the ferromagnetic Ising model near the Curie temperature due to critical
slowing down. This problem is particularly acute for the square lattice
dimer model, which has critical correlations. Elementary moves with
many participating degrees of freedom can give rise to a substantial
reduction of the Monte Carlo correlation time.

Several algorithms have been proposed to locate long connected transition
graphs.\cite{barkema} Typical problems in this context include diverging
overhead (most of the time spent looking for a long loop is essentially
wasted) and diminishing return (e.g. a long loop may end up invading
the full system, or dynamically important moves may not be generated
for sufficiently large systems).  The `pocket algorithm' we present here
for hardcore dimers generates subgraphs from two configurations related
by a global lattice symmetry.  The algorithm obtains long transition
graphs with minimal overhead: on lattices of size $L \times L$, no moves
need to be rejected, and any dimer encountered during the construction
participates in the move. In addition, there is no need for bookkeeping.

\begin{figure}
\centerline{ \psfig{figure=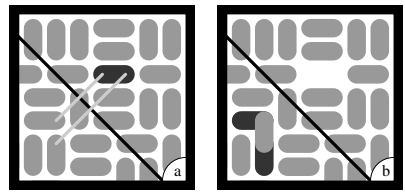,height=2.5cm} }
\caption{
Dimer algorithm. a: Symmetry axis and seed dimer (dark gray),
sole initial element of the set $\mathcal{P}$, the `pocket'. b: An
element of the pocket has been reflected with respect to the symmetry
axis (and transferred from $\mathcal{P}$ to $\mathcal{A}$, the set of
regular dimers). It overlaps with two (dark) dimers, which are transferred
from $\mathcal{A}$  to $\mathcal{P}$.  The move construction comes to an
end when $\mathcal{P}$  is empty.  }
\label{f:dimer_algo}
\end{figure}

The pocket algorithm proceeds in a series of moves, at the beginning
of which we randomly pick a symmetry axis, and a seed dimer (see
Fig.~\ref{f:dimer_algo}). At this initial stage of the move
construction, the seed is the sole element of a set $\mathcal{P}$ (the
`pocket', filled with dark dimers), while all other (light gray)
dimers belong to a set $\mathcal{A}$. Further in the move, dimers are
shuffled around between sets $\mathcal{A}$ and $\mathcal{P}$ in the
following way: At each step, an arbitrary element $i$ of $\mathcal{P}$
is reflected with respect to the symmetry axis, and moved to
$\mathcal{A}$. If $i$ overlaps with other elements of $\mathcal{A}$
(dark dimers shown in Fig.~\ref{f:dimer_algo}b), the latter are moved
from $\mathcal{A}$ to $\mathcal{P}$. The move is completed as soon as
the pocket $\mathcal{P}$ is empty. During the whole construction, the
algorithm keeps no memory of the graph, and final success is
assured by the underlying symmetry of the lattice in combination with
the hardcore property of the dimers.

The pocket algorithm is equivalent to the pivot-cluster
algorithm,\cite{Dress} which has been applied successfully to several
hard-sphere problems.\cite{binary,glass} It provides a greatly simplified
implementation, as the Monte Carlo move is constructed without ever
explicitly considering the cluster.  In the specific application of dimer
models, we also use reflections with respect to symmetry axes rather than
point reflections. On the square lattice, for example, the algorithm
is ergodic if we allow both diagonal and horizontal-vertical axes. The
first choice allows to change the numbers of horizontal/vertical dimers,
the second one permits to move through the different winding number
sectors. The algorithm was tested against exact enumeration
for systems up to size $L=6$.

Transition graphs generated  by the pocket algorithm, as shown in
Fig.~\ref{f:graph_and_loop}, are symmetric with respect to the symmetry
axis, and cross it at most twice.

An analogous property holds in higher dimensions. For dimers on a
three-dimensional lattice, for example, the appropriate operations
are reflections with respect to randomly chosen symmetry planes of the
(finite) lattice. Again, any transition graph is symmetric with respect
to this symmetry plane, and crosses it at most twice. This places a
strong constraint on the transition graph, which cannot invade space.
If we used reflection with respect to a point, as was appropriate in
other applications,\cite{Dress} the move would conserve the total
number of dimers of a given orientation, and also generate
frequent transition graphs completely invading the lattice.

Transition loops generated by the algorithm allow to flip not only
the loop itself but also all the dimers inside the (gray) loop area,
which is symmetric with respect to the axis (see the example in
Fig.~\ref{f:graph_and_loop}).  This modification, as well as simple
generalizations where the pocket is initially populated by more than one
dimer, or by a specially positioned dimer,  have not yet been studied.
We also note that this algorithm applies to other models that can be
mapped onto (not necessarily nearest-neighbor) 
dimer models.\cite{FishIsing}

For the undoped system on the square lattice, the pocket algorithm
generates transition loops of average length proportional to $L $.
A more detailed analysis is provided in the appendix.

\section{Dimer models at finite monomer density}

We now turn to the problem of the correlations of monomers in a
background of dimers on the square and triangular lattices. We
evaluate the monomer two-point correlation function for a range of
dopings. 

The monomer two-point function is defined as follows. Let $Z(r)$\ be the
partition function of the monomer-dimer problem with a monomer fixed at
the origin and another at site $r$: $Z(r)$\ counts the number of allowed
configurations subject to the pair of monomers being fixed. The two-point
function is then defined to be proportional to $Z(r)$: $C(r)\equiv
Z(r)/Z_0$; the constant of proportionality, $Z_0$, is somewhat arbitrary.
We choose to normalize the correlations such that a random distribution
of monomers would give $C(r)=1$.  For the two-monomer problem however,
to make contact with the previous literature,\cite{Fisher63} we normalize
the pair correlations such that (a) $C(1)=1/z$ and (b) list $z C(r)$
in Tab.~\ref{tab:numvsana}, where $z$\ is the coordination of the lattice.

For the square lattice, the two-monomer correlations were obtained by
Fisher and Stephenson\cite{Fisher63} and by Hartwig\cite{Hartwig} by
perturbing the Pfaffian matrices introduced by Kasteleyn.\cite{kasteleyn}
For the triangular lattice, there is as yet no analytic expression
for the asymptotics of the monomer correlations,\cite{fn-ana}  and we
have obtained the exact $L=6$ results by  explicit enumeration.

\subsection{The square lattice}

\begin{figure}
\centerline{ \psfig{figure=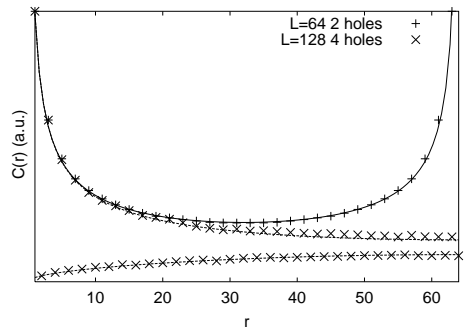,height=4.5cm } }
\caption{Monomer correlations along a coordinate axis: (1) pair
correlations on the square lattice with $L=64$. The fit is to an
analytic asymptotic expression generalized to include periodic boundary
conditions. Same-sublattice correlations vanish. (2) Four monomers on a
square lattice with $L=128$. The lower curve corresponds to monomers on
the same sublattice, the upper to monomers on different sublattices. The
upper curve is again compared to the appropriate asymptotic function,
whereas the lower curve is fitted to a power-law $r^{1/2}$. The deviation
of the upper curve from the asymptotic expression for large $r$ is due
to screening.}
\label{fig:ana2_4hole}
\end{figure}

To test the algorithm, first consider two-monomer correlations,
calculated by several authors for the square lattice.  In
Fig.~\ref{fig:ana2_4hole} and Tab.~\ref{tab:numvsana}, we compare our
data against the analytical results. Note that the fit against the
analytical asymptotic expression in Fig.~\ref{fig:ana2_4hole} was done
using a distance variable $\tilde{r}=\sin{(\pi r/L)}/(\pi/L)$\ to take
into account periodic boundary conditions for a torus of length
$L$.\cite{fn-per}

\begin{table}
\begin{center}
\begin{tabular}{|c||c|c||c|c|c|}
\hline
$r$ & exact\cite{Fisher63} & MC $L=64$& exact $L=6$& MC $L=6$&MC  $L=12$   \\
\hline \hline
1 & 1      &1       &1       &1       &1  \\ 
2 & 0.6366 &0.638  &0.8792  &0.8789  &0.8792   \\ 
3 & 0.5947 &0.595  &0.8984  &0.8980  &0.8976   \\ 
4 & 0.4821 &0.484  &  &  		&0.8969   \\ 
5 & 0.4506 &0.452  &  &  		&0.8970   \\ 
6 & 0.3995 &0.402  &  &  		&0.8966   \\ 
\hline
\end {tabular}
\end{center}
\caption{ Monte Carlo versus exact results for monomer pair correlations,
$zC(r)$, on the square (left) and triangular (right) lattices.
The normalization here is $zC(1)=1$. Even distances in the square
lattice refer to sites on the adjacent row.\cite{Fisher63} For the
simulations presented here, the statistical error is about $10^{-3}$
for both lattices.
}
\label{tab:numvsana}
\end{table}

The fit is against the analytic asymptotic expression, and
in Tab.~\ref{tab:numvsana}, we show a comparison of analytical and
numerical results.\cite{footspecial} The results agree to within the numerical accuracy
of $10^{-3}$, which improves even further for larger monomer densities,
$\rho$, as  the signal-to-noise ratio increases with the number of
monomer pairs as $\rho^2$. A more accurate estimate could be obtained
by finite-size scaling.

For two monomers on opposite sublattices, the form of the correlations,
$C(r)=a_- r^{-1/2}$\ defines an entropic Coulomb interaction between
the two monomers: $V(r)=-\frac{1}{2}\log r-\log a_-$. Note that whereas
the pair of monomers are critically confined -- $C(r)\rightarrow 0$\
algebraically as $r\rightarrow\infty$ -- the expectation value of their
separation grows with system size: $\langle r \rangle \propto L$.

As the square lattice is bipartite, dimers have one end on each
sublattice. This means that on a torus, there are no dimer configurations
with the two monomers present on the same sublattice: in this case,
$C(r)\equiv 0$.

The question whether there is an effective interaction of the same form
is nonetheless interesting. To avoid the sublattice constraint, we have
investigated \emph{four} monomers on a large lattice of size  $L=128$.
For monomers on opposite sublattices, the results are essentially
unaltered from the two monomer case, at least for separations which are
not too large ($r\ll L$); for monomers on the same sublattice, we now
obtain $C(r)=a_+ r^{+1/2}$, so that the monomers on each sublattice act
as if they had equal and opposite Coulomb charge.

The $L$-dependence of the ratio $a_-/a_+$\ is fixed by the fact that
the square lattice is at a critical point. This allows to write the
finite-size scaling ansatz $C(r)= \alpha L^{-2} c(r/L)$, where $c(x)$\
is a function only depending on the {\em ratio} of $r$ and $L$, and
$\alpha$\ does not depend on $L$. The prefactor $L^{-2}$\ follows from
the requirement that the area underneath the curves remain fixed as
$L\rightarrow\infty$. With $c_\pm(x)\propto x^{\pm 1/2}$, we obtain
$a_-/a_+\propto L$. A fit for $L=128$ and four holes gives $a_-/a_+
\approx L/2$.

The dimers thus generate a Coulomb interaction for the monomers, $V(r)=(q
q^\prime/2) \log r$, with a sublattice-dependent  sign of the monomer
charge $q$ ($|q|=1$).  It is natural to  expect this Coulomb interaction
to be screened if the monomer density is finite, and their correlations
to be no longer critical. In fact, for a finite density,  $\rho$, of
monomers, a new length scale arises, namely the average interparticle
spacing, $d \sim \rho^{-1/2}$.

To analyze the details of the screening analytically, we model the
monomer system by a classical, continuous two-dimensional
two-component plasma. With a dimensionless coupling of $\beta=1/2$
(the prefactor of the logarithm), this plasma is located well in its
high-temperature phase. This phase extends up to $\beta=2$, where the
continuum approximation breaks down as the plasma collapses in the
absence of a repulsive hard core to the interaction.

In this high-temperature regime, we expect Debye screening to be
operative. The resulting long-distance connected correlations have
been obtained in Ref.~\onlinecite{samaj} 
(where an expression going beyond
Debye screening is given); they are 
\bea  
C(r) = 1  \pm
\chi\sqrt{\frac{\pi\lambda_D}{2r}}\exp(-r/\lambda_D),
\nonumber 
\eea 
with the Debye screening length: $\lambda_D=1/\sqrt{2 \pi \beta \rho}$,
and with $\chi=\beta$\ to lowest order in the coupling. The choice of
sign depends on the sublattice.

With the new length scale, $\lambda_D\propto d$, the criticality of the
correlations is destroyed, although the critical behavior remains visible
for $r\ll \lambda_D$. This is why it was possible to find the two-particle
interaction from the four-monomer problem above.

\begin{figure}
\centerline{ \psfig{figure=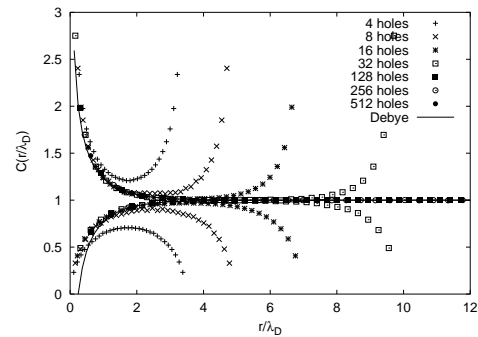,height=4.5cm } }
\caption{Monomer correlations $C$ for different monomer densities, 
against the scaled distance
variable $r/\lambda_D$, for the square lattice with  $L=64$.
}
\label{fig:ana64scale}
\end{figure}

In order to determine whether this scenario is indeed realized for
the monomer-dimer mixture, we have plotted $C(r)$\ versus the scaled
distance $r/\lambda_D$ for varying monomer densities for a system of size
$L=64$\ (Fig.~\ref{fig:ana64scale}).

From the form of the expression for $C(r)$, all curves should collapse
on top of one another. However, for a finite system, one has to take
into account the presence of two length scales, $L$ and $\lambda_D$,
with the added feature that $\lambda_D<L$, so that a simple one-parameter
finite-size scaling ansatz no longer applies.
For $\lambda_D\ll L$ (large monomer densities), the collapse
works indeed very well.

\begin{figure}
\centerline{ \psfig{figure=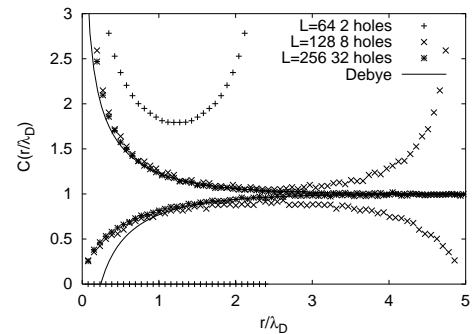,height=4.5cm } }
\caption{Monomer correlations $C$ 
against the scaled distance variable $r/\lambda_D$ at fixed monomer density 
$\rho = 2/64^2$ for square lattices of size $L=64,\ 128,\ 256$.}
\label{fig:anafixedrho}
\end{figure}

The  data do not  collapse at small densities, which is  a finite-size
effect: In Fig.~\ref{fig:anafixedrho}, we plot  the monomer correlations
at a fixed monomer density $\rho=2/64^2$\ for $L=64,\ 128$ and 256.
Whereas the two-hole problem suffers from the sublattice pathology,
one can see that the data for $L=128$ and 256 collapse nicely for short
distances, with disagreement arising for distances around $L/2$ where
the periodic boundary conditions become important.

In the region where they collapse, the curves in
Figs.~\ref{fig:ana64scale} and \ref{fig:anafixedrho} agree with the
asymptotic Debye prediction for large values of $r/\lambda_D$.  At small
$r$, the effectively unscreened behavior is visible, leading again to
a unique curve for $C(r)$.

\subsection{The triangular lattice} 

Dimer correlations on the triangular lattice are known to be
short-ranged,\cite{MStrirvb} and not critical as is the case for the
square lattice.\cite{Fisher63} This leads one to expect that the monomer
correlations will also be short-ranged as the presence of a monomer
should have little effect at a distance large compared to the dimer
correlation length.

\begin{figure}
\centerline{ \psfig{figure=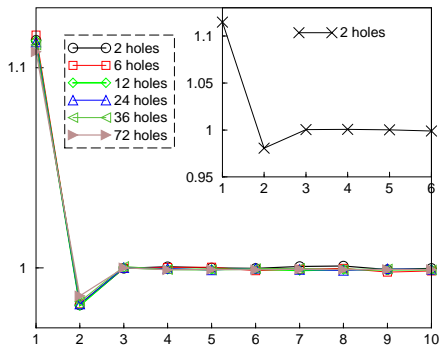,height=4.5cm } }
\caption{Monomer correlations $C$,  along a coordinate axis at
variable doping for a lattice of size $L=24$. The abscissa is
in units of lattice spacings. Correlations are normalized against a
random distribution. Inset: two-monomer correlation in a lattice of size
$L=12$, In both cases, the $x$-axis denotes separation along one
of the lattice vectors. Note that the $y$-axis
only varies by less than 15\%.}
\label{fig:corrtri}
\end{figure}

We have plotted the correlations of a pair of monomers in the inset to
Fig.~\ref{fig:corrtri}. The correlation length indeed is so small that
already for a triangular lattice of size  $L=12$, finite-size
effects are no longer detectable. This is also shown in
Tab.~\ref{tab:numvsana}.

At `large' distances, the correlations tend to a constant, nonzero value
of $C(\infty)=0.1495\pm0.0001$. (This value is given, in keeping with the
convention of the square lattice case, with a normalization such that
$C(1)=1/6$.)  They do so with an extremely short correlation length,
which we cannot determine accurately as there are only about 3 data
points above the noise level. This makes it hard to fit an asymptotic
form, as the presence of a power-law and/or oscillatory multiplicative
factor cannot be detected on this basis.  From the absence of any signal
for distances beyond $r=3$, we conclude that the correlation length is
less than one lattice spacing.

This nonzero large-distance limit of the monomer correlation function
implies that monomers in the classical hardcore dimer model on the
triangular lattice are deconfined -- the free energy for monomers a
distance $r$\ apart does not diverge as $r\rightarrow\infty$. This
contrasts with the square lattice, where this quantity diverges
logarithmically.

Upon further doping, the correlations barely change (see
Fig.~\ref{fig:corrtri}); a small decrease of the correlations only
becomes visible for a doping level of $1/8$. This is not unexpected as
the average monomer spacing always remains well above the two-monomer
correlation length.

\section{Conclusion}

We have studied the correlations of monomer-dimer mixtures on the square
and triangular lattices with a new Monte Carlo algorithm.  The monomers
interact via the background of dimers: the critical correlations of the
dimers on the square lattice give rise to a two-monomer interaction of
long-range Coulomb nature, whereas the disordered triangular correlations
generate an exponentially decaying one.

The triangular model is deep in its disordered phase, where it remains in
the presence of a finite density of monomers.  The square model, located
at a phase boundary, is driven into a disordered phase on doping.
This effect is rendered analytically by modeling the monomers as a
two-dimensional two-component plasma, with the long-range correlations 
given by a simple Debye-screened form. 

\section*{Appendix}

\begin{figure}
\centerline{ \psfig{figure=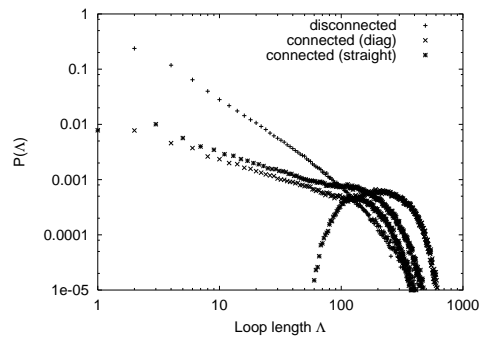,height=4.5cm } }
\caption{
Distribution of transition loop lengths for the square lattice with
$L=64$.
}
\label{f:looplength_sq_64}
\end{figure}

In this appendix, we analyze the distribution of transition graph
lengths in the undoped dimer model, where only loops are generated.
These loops can be of several types: disconnected pairs of loops
not touching the symmetry axis, connected loops passing through the
symmetry axis, but not winding around the box, and finally those winding
around the box. Each of these distributions strongly depends on the
lattice and on the details of the choice of axes.  The distribution
of loop lengths, $\Lambda$, generated by the pocket algorithm is
shown in Fig.~\ref{f:looplength_sq_64} for the square lattice.
\begin{figure}
\centerline{ \psfig{figure=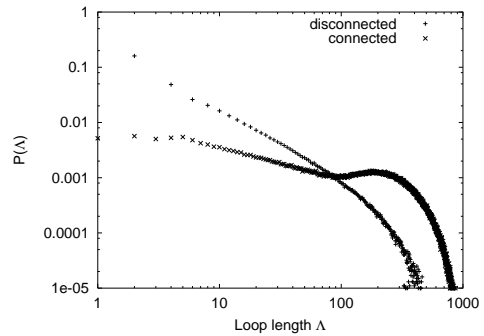,height=4.5cm } }
\caption{
Distribution of transition loop lengths for the triangular lattice with
$L=64$.
}
\label{f:looplength_tri_64}
\end{figure}
Whereas the disconnected loops follow a nice power law distribution
$P(\Lambda)_{\text{disconnect}} \sim \Lambda^{-1.5}$ for $\Lambda < L$,
and are always made up of an even number  of dimers, the connected loops
are  distributed as $P(\Lambda)_{\text{connect}} \sim \Lambda^{-0.75}$.
For the reflection about a diagonal,  the loop lengths are always even,
whereas they are odd for the horizontal-vertical reflections that do
not wind around the lattice.  If a loop does wind around the box, its
length can be odd or even.  Notwithstanding this intricate structure,
we find that the mean loop length grows as $ L^{\mu}$, with $\mu=1.00
\pm 0.01$.
On the triangular lattice, the distribution of loop lengths is also 
nicely split into connected and disconnected loops 
(see Fig.~\ref{f:looplength_tri_64}). There, we find for the 
mean loop length 
$\mu=1.47\pm0.02$.
We note that the algorithm is biased towards longer
subgraphs of the transition graph of a configuration with its reflection
with a factor proportional to their length.

Finally, it would be interesting to analyze the exponents in more detail,
in particular as the square model maps onto a height model,\cite{Kondev}
whereas the triangular one does not.\cite{Ising}

\section*{Acknowledgements}
RM would like to thank John Chalker, Paul Fendley, David Huse, Jane
Kondev and Shiavji Sondhi for useful discussions.


\begin{thebibliography}{99}
\bibitem{footspecial} For these simulations, the initial seed of the 
pocket algorithm was chosen to be one of the monomers.
\bibitem{fn-per} By conformal invariance, this form is strictly
applicable only for an infinite strip of width $L$ for a system with
critical correlations.
\end{thebibliography}
\end{document}